\documentclass[12pt]{iopart}
\usepackage{iopams}
\usepackage{lcy}

\newcommand{\dfr}[2]{\frac {\displaystyle #1}{\displaystyle #2}}

\begin{document}

\title[A description of $\alpha $- and $\beta $-relaxation supercooled in liquids]
{A description of $\alpha $- and $\beta $-relaxation in supercooled liquids within the framework of the continuous
theory of defects }
\author{M G Vasin}
\address{Physical-Technical Institute of the Ural Branch of the Russian Academy of Sciences,
 132 Kirov Street, Izhevsk, 426000, Russia}
\ead{vasin@udm.net}
\begin{abstract}
Supercooled liquid is described in terms of the theory of defected states of bond orientational order using the
continuous theory of defects. It is shown that the initiation of a local orientation order leads to the breakdown of
the initial SO(3) symmetry of the theory, as a result the plastic deformation fields of the linear defects become
Abelian. The expressions for linear dislocation and disclination tensor potentials are derived within the framework of
the quasistationary adiabatic approximation. It is shown that $\alpha $-relaxation and $\beta $-relaxation of
supercooled liquid can be described as relaxation processes in the disclination and dislocation subsystems accordingly.
Disclination topological moment is proposed to use as an order parameter at the description of glass transition.
\end{abstract}
 \submitto{\JPCM} \maketitle

\section{Introduction}
It is known that the translation symmetry absence in liquids (does not entail) (is not accompanied by) an absolutely
random atom order. Most liquids under low temperatures possess a short-range or middle-range order. This ordering is
determined by the directions and lengths of valence bonds, valence angles, extents of the atom and ionic radiuses, etc.
In contemporary journals this order is called ''local order" \cite{Pat}, ''locally favored structures"~\cite{Tanaka} or
''bond orientational order"~\cite{4}, while the liquid state is called Frenkcel phase \cite{Kov}, high density phase
\cite{Stanley} and so on. The pioneering liquid state theory, based on the bond orientational order conception, was
suggested by Frenkcel \cite{1}. In spite of the objective complications of experimental investigations and theoretical
description of liquids there is appreciable progress in development of this conception. The plenty experimental
results, which  sustain this conception, (see \cite{Kov} for example) and theoretical works, which develop one,
\cite{3} have recently appeared. Among of them let us note the model of defected states of bond orientational order,
suggested by Nelson \cite{4} with the purpose of a disordered molecular systems description. Our interest to this
approach is caused by the opportunity to use it for description of the supercooled liquids and the glass transitions
\cite{Rivier}.

The general idea of the considered model can be formulated in the following way: Under the high temperatures a liquid
is an isotropic (quasi-gas) state of the condensed matter, which is characterized by absence of any short-range atom
ordering and, as a result, locally SO(3)-symmetric. Under the temperature decrease or the pressure increase the initial
system ordering happens. In this case the short-range order, which is locally symmetric relative to some point group,
forms. Since this symmetry depends only on atom interaction, it is natural to assume that in most cases they are not
among the Fedoroff groups, therefore, the ground state of these systems is a noncrystalline (a long-range order is
absent) and infinite degenerated one owing to the geometric frustrations.

The topological defects (dislocations and disclinations), which correspond to these frustrations, are basic structure
elements of the system. It is important to emphasize that in contrast to the crystal defect, which increase the system
energy, the defects in the considered model are essential structure elements, which compact the pacing and make the
disordered system more energy-optimal than the crystal one~\cite{4}. In order to destroy the topology defects great
energy consumptions are necessary. However, because of the frustrations and the thermal flotations the system of
topology defects is moving. This motion is a cooperative atom motion, and due to this the condensed matter maintains a
fluidity.

\section{The gauge field theory Lagrangian}
The gauge defect theory is a method of the disordered molecular system description, which is based on the mathematics
of the gauge field physics. This approach implies a fundamental role of the defects in the properties formation of the
disordered molecular systems. Its basis was developed in the works of Lihkachev \cite{5}, Kadich and Edilen \cite{6},
Nelson \cite{4}, etc.

In the most fundamental form, which was suggested in \cite{6}, the gauge defect theory is represented within the
generalized theory of the Yang--Mills gauge fields. To write down the Lagrangian of the defect system, let us make use
of the standard theory of elasticity. The simplest Lagrangian, describing the system with elastic deformations, has the
following form:
$$
  L_0=\dfr 12\rho _0\partial _4\chi _i\partial
  _4\chi _i-\dfr 18 \left[ \lambda \,(u_{ii})^2
  +2\mu \,u_{ij}u_{ij}\right]\,,
$$
where $\chi _i(\bar r,\,t)$ is the elastic strain field, $\lambda $ and $\mu $ are the Lame constants (the elasticity
and shear modulus accordingly), $\rho _0$ is the mass density (which is considered to be constant for simplicity), and
$u_{ij}$ are the relative deformation components:
$$
  u_{ij}=C_{ij}-\delta _{ij}=\partial _i\chi _l\partial
  _j\chi_l-\delta _{ij}
$$
(Roman letters $a,\,b,\,\ldots $ are used to denote the space components set $\{ i \}=\{1,\,2,\,3\}$, and Greek letters
$\alpha ,\,\beta ,\,\ldots $ are used to denote the full index set, including the time component $X^4$, $\{\alpha
\}=\{1,\,2,\,3,\,4\}$).

According to the gauge theory of dislocations and disclinations, the plastic deformation of the matter structure can be
considered as a breakdown of the homogeneity of the rotation and translation (SO(3)$\triangleright $T(3)) groups
action. In order to take into account these homogeneity breakdowns, the compensating fields ($A_a^{\alpha }$ and
$\varphi ^i_b$) are introduced into the Lagrangian, and the transition from ordinary to covariant derivatives is
effected:
$$
  \partial _{\alpha } \chi _i\to B_{\alpha i}=\partial _{\alpha }\chi _i+\varepsilon
_{ijl}A_{\alpha j}\chi _l+\varphi _{\alpha i},
$$
where $\varepsilon _{ijl}$ are three generating matrixes of the semisimple group SO(3). After that, the Lagrangian
$L_0$ is replaced by the new Lagrangian
$$
  L=L_0+s_{\varphi }L_{\varphi }+s_AL_A,
$$
where $s_{\varphi }$ and $s_A$ are free parameters of the theory, and the first term describes elastic properties of
matter:
$$
\displaystyle L_0=  \dfr 12\rho _0B_{i4}B_{i4}-\dfr 18 \left[ \lambda (E_{ii})^2+2\mu E_{ij}E_{ij}\right] ,
$$
where $E_{ij}=B_{li}B_{lj}-\delta _{ij}$ is the unit strain tensor. The second term,
$$
  \displaystyle s_{\varphi }L_{\varphi }=-\dfr 12 s_{\varphi }D_{i\alpha \beta }D_{i\alpha \beta },
$$
describes the dislocations; the following notation is used here:
$$
  \begin{array}{c}
  \displaystyle D_{i\alpha \beta }=\partial _{\alpha }\varphi _{i\beta }-\partial _{\beta }\varphi
  _{i\alpha }+\varepsilon _{ilj}\left( A_{l\alpha }\varphi _{j\beta }
  -A_{l\beta }\varphi _{j\alpha }+F_{l\alpha \beta }\chi _j\right),
  \\[12pt]
  \displaystyle F_{i\alpha\beta }=\partial _{\alpha }A_{i\beta }-\partial _{\beta }A_{i\alpha }+
  \varepsilon _{ijl}A_{j\alpha }A_{l\beta },
  \end{array}
$$
The third term,
\begin{equation}\label{1}
  s_AL_A=-\dfr 12s_AF_{ijl}F_{ijl}+ \dfr 1{2\zeta}s_AF_{i44}F_{i44},
\end{equation}
describes disclintions. The Yang-Mills fields, $A_{i\alpha }$ and $\varphi _{i\alpha }$, describe disclinations and
dislocations accordingly.

The gauge defects theory is a general one, since it considers the case of the ideal isotropic matter, which is locally
symmetrical with respect to the SO(3)-group of rotations in the three-dimensional space. This theory is non-Abelian one
\cite{3, 6}, which is the main problem of the considered approach. However, this paper will argue the idea that in the
systems, which have a bond orientational order, this trouble disappears.

\section{The Lagrangian symmetry breakdown}
Note, that a molecular system can be SO(3)-symmetrical just as Hamiltonian only at the relatively high temperatures. At
low temperatures, according to the considered model, a local bond orientational order appears there. Defects can exist
only in local ordered systems. Then the system's choice of a ground state and appearance of an order parameter, ${\bf
Q}(\vec r)$, with some internal state space, $M^n$, lead to the spontaneous symmetry breakdown. As a result the ground
state symmetry does not agree with the Hamiltonian's SO(3)-symmetry.

The broken symmetry is the bond orientational order symmetry, which is generally determined by the nature of the
chemical molecular bonds. Let us consider two important cases: 1) In the systems with the spherical interaction
potential the tetrahedral pacing is the most preferred one, which depends the icosahedral form of the short range
ordering. It is known that the icosahedral symmetry group, Y, is isomorphous to the finite group of the proper
rotations of ${\mathbb{R}}^3$ space, which has three orbits (see appendix 1). Each orbit correspond to the rotation a
round one of the symmetry axes. Since each of these orbits coincides with the corresponding internal state space, one
can assert that the symmetry groups of these internal states are point symmetry subgroups in the SO(3). 2) Director is
the order parameter in the nematic liquid. The ground state of this system can be arbitrary configuration of the
codirectional directors. The system's <<choice>> of one of the ground states leads to the SO(3)$\to $SO(2) symmetry
breakdown, as a result the system remains symmetrical in relation to rotations around the selected axis.

Thus in the polytetrahedral and monoaxial nematic models the symmetry groups, which keep the order parameter, are
subgroups of the SO(2). The nonlinear over $A_{ij}$ part of disclination field tensor vanishes, and the fields of the
linear topological defects become Abelian:
$$
F_{i\alpha\beta }=\partial_{\alpha }A_{i\beta }-\partial_{\beta }A_{i\alpha }+\varepsilon_{ijl}A_{j\alpha }A_{l\beta }
\qquad \to \qquad F_{i\alpha\beta }=\partial_{\alpha }A_{i\beta }-\partial_{\beta }A_{i\alpha }.
$$
This leads to a significant simplification of the considered systems analysis, since the Lagrangian is out of nonlinear
$A_{ij}$ terms, which are typical for the Yang--Meals theory.

\section{ The linear topology defects fields in the quasistationary adiabatic approximation}

In order to find the tensor potentials of the linear defects one needs to do some simplifications: First of all let us
describe the defects system in the quasistationary approximation ($\partial _4A_{i\alpha }=0$). In case of slow
relaxation this seems to be appropriable. Besides, in this paper we will be interested in the typical for amorphous
systems low-angled disclinations with $\nu \sim 0.2$ Frank index. In this case the considered model can be linearized:
\begin{equation}\label{f}
\begin{array}{rl}
&\chi _i=\delta _{ia}x_a+\varepsilon u^{(1)}_{i}+\varepsilon ^{2}u^{(2)}_{i}
+\ldots ,\\[12pt]
&A_{\alpha i}=\varepsilon A_{\alpha i}+\varepsilon ^2A^{(1)}_{\alpha i}+ \ldots ,\\[12pt]
&\varphi_{\alpha i}=\varepsilon \varphi_{\alpha i}+\varepsilon^2 \varphi^{(1)}_{\alpha i}+\ldots ,
\end{array}
\end{equation}
where $\varepsilon \sim \nu $ is the small parameter, $u_i=\varepsilon u^{(1)}_{i}+\varepsilon ^2u^{(2)}_{i}+\ldots $
is the displacement vector. For simplicity let us assume that the free topology defects do not interact with each
other. Below we will show that all these simplifications are equivalent to the adiabatic approximation. This seems to
be appropriable in view of the significant difference between the dislocation and cisclination subsystem relaxation
times.

In order to determine the tensor potential of the linear disclination segment let us minimize the disclination term of
the Lagrangian,
$$
L_{A}=\dfr 14F_{\alpha ij}F_{\alpha ij}.
$$
In \cite{Vasin} it was shown that the quasistationary solution for the gauge field tensor potential of the linear
disclination segment, $dl_{\alpha }$, has the form of
\begin{equation}\label{2}
A_{i 4}=x_{\alpha }\left( \dfr C{r^3}+2C_1\right) dl_{i}, \qquad A_{i k}= \, \varepsilon _{i kj}x_j\left( \dfr
C{r^3}+C_1 \right)dl_{i }.
\end{equation}
It is significant that in this expression there is no a summation over $\alpha $, and $C_1=0$ according to $A_k^{\alpha
}(\infty )=0$.

Then from (\ref{1}) one can determine the linear dislocation field. In the linear approximation the $A\varphi $ product
is proportional to $\varepsilon ^2$, therefore one can neglect the first two terms in the brackets:
$$
D_{i\alpha \beta }=\partial _{\alpha }\varphi _{i\beta }-\partial _{\beta }\varphi
  _{i\alpha }+\varepsilon _{ilj}F_{l\alpha \beta }\chi _j,
$$
where
$$
F_{i\alpha\beta }=\partial _{\alpha }A_{i\beta }-\partial _{\beta }A_{i\alpha }.
$$
As it was noted above the linear approximation leads to the neglect of the dislocation birth--annihilation processes on
the disclinations.

The dislocation and disclination systems have intricate configurations. Therefore, in spite of the above
approximations, the problem of the theoretical description of such systems is a difficult task. But it can be
significantly simplified when the high defect mobility is considered. In case of high defect concentration they screen
the elastic energy of each other by the formation of the disclination and dislocation loops. The screening leads to the
minimization of summary local topology charge. This minimizes the dislocation and disclination subsystems energy, since
the elastic energy density of the defect loop decreases faster than one of the single linear defect. In case of
disclinations it is $1/r^3$ against $1/r$. As a result the tensor potential of the disclination subsystem field is
determined by the topology disclination moment, $Q_{ij}(r)$: $A_{ij}=\dfr {Q_{ij}}{r^3}$ (see appendix 2). Substituting
this potential form to the above expression one can write
$$
\begin{array}{c}
\displaystyle D_{i\alpha \beta }=\partial_{\alpha }\varphi_{i\beta }-\partial_{\beta }\varphi_{i\alpha
}+\int\varepsilon_{ilj}\left(\partial_{\alpha }A_{l\beta }(\bar x-\bar r)-\partial_{\beta }A_{l\alpha }(\bar x-\bar
r)\right)x^j dr=
\partial_{\alpha }\varphi_{i\beta }-\partial_{\beta }\varphi_{i\alpha }.
\end{array}
$$
It follows that the linear dislocation field potential has the same form like the disclination one:
$$
\varphi_{ij}=\varepsilon_{ijk}\dfr{x_k}{x^3}dl_i.
$$

Then one can get the tensor potentials of arbitrary configurations of the dislocation and disclination fields. Taking
into account the screening the interaction energy of the linear defects system can be represented in the form of the
interaction energy of the local topology moment system, and the problem is reduced to the known one.

\section{Topology moment interaction}

In spite of the fact that the linear dislocation and disclination tensor potentials have the similar form, they have
different contributions in the elastic interaction. This determines the significant difference between the disclination
and dislocation subsystem relaxation dynamics.

The stress field is produced by the dislocations and disclinations.  Keeping only nondiagonal elements of the
deformation tensor in the first infinitesimal order one can write it in the form
$$
\sigma _{ai}\simeq\mu E_{ai}=\mu \left( B_{aj}B_{ji}-\delta _{ai}\right)=\mu \left[ \varepsilon _{\alpha
al}x_lA^{\alpha }_i+\varepsilon _{\alpha il}x_lA^{\alpha }_a+\varphi_{ai}+\varphi_{ia}+\partial _i u_a+\partial
_au_i\right] .
$$
The elastic interaction energy,
$$
U^{E}=\int E_{ai}\sigma _{ai} dV=U^R+U^T+U^{RT},
$$
includes the inter-dislocation, $U^T$, and inter-disclination, $U^R$, interaction terms, as well as the
dislocation--dislocation interaction $U^{RT}$ term. In spite of the fact that the last term is important in the
calculations, in order to simplify the qualitative analysis we will limit ourselves only with the first two terms.

As it was noted above, the elastic interaction energy of the disclination network can be represented in the form of the
interaction energy of the topology moments system:
$$
U^R=\dfr {\mu }2\int \varepsilon _{\alpha al}x_lA^{\alpha }_i \varepsilon _{\gamma aj}x_j'A'^{\gamma }_i dV\approx\dfr
{a^2}2\sum\limits_{\alpha ,\,\beta } Q^{R}_{il}(\vec r_{\alpha })J^R_{ijlk}(\vec r_{\alpha \beta }) Q^R_{jk}(\vec
r_{\beta }),
$$
where $\alpha $ and $\beta $ label the moments, $a$ is the typical linear dimension of disclination loops, and
$$
   J^R_{lkij}(\vec r)\approx {\mu }\dfr {\delta _{ij}}{2|\vec r|^3}{\left(
   \delta ^{lk}-3\dfr {r^{l}r^{k}}{r^2}\right)} .
$$
The elastic energy of the dislocation topological moments system has the form of
$$
\displaystyle U^T=\dfr {\mu }2\int\varphi_{ai}\varphi'_{ai}dV\approx\dfr {a^2}2\sum\limits_{\alpha ,\,\beta }
Q^{T}_{il}(\vec r_{\alpha })J^{T}_{ijlk}(\vec r_{\alpha \beta }) Q^{T}_{jk}(\vec r_{\beta }),
$$
where
$$
   J^{T}_{lkij}(\vec r)\approx {\mu }\dfr {\delta _{ij}}{2|\vec r|^5}{\left(
   \delta ^{lk}-3\dfr {r^{l}r^{k}}{r^2}\right)} .
$$
Comparison of the expressions for the disclination interaction and dislocation interaction makes it clear that the
first one is the long-range interaction
$$
J^R(r)\sim 1\left/r^3\right.,
$$
while the second one is short-range:
$$
J^T(r)\sim 1\left/r^5\right.,\quad 5>d=3.
$$
This is a very important difference, since it leads to the difference of relaxation dynamics of the disclination and
dislocation subsystems, which is the cause of the experimentally observed decoupling between rotational and
translational diffusion \cite{9}.

In order to display it let us represent the energy functional of the disclination subsystem in the form of the
functional of the frustration limited domains theory \cite{3, Kiv1}:
$$
U^R=\int \left\{ \dfr 12 |\partial_{\mu }{\bf Q}^R(\vec x)|^2+\dfr 12\,\tau |{\bf Q}^R(\vec x)|^2\right\}d^3x-\dfr 12
\int\int {\bf Q}^R(\vec x){\bf J}^R(|\vec x-\vec y|){\bf Q}^R(\vec y) d^3xd^3y,
$$
where ${\bf Q}$ is some local structural variable (order parameter). The first item is a short-range interaction caused
by continuity of ${\bf Q}$, the second item is a weak long-range interaction, which describes the disclination moments
interaction. In \cite{Kiv1} it was shown, that presence of a long-range interaction in such system leads to its
frustration. It allows us to assert, that because of the long-range interaction the disclination subsystem is
geometrically frustrated. The system's frustration leads to the slow non-Arrenius dynamics of relaxation processes
proceeding in this system~\cite{3, Vas}. Thus in the considered system the relaxation of the disclination subsystem is
$\alpha $-relaxation.

On the other hand the short-range interaction of the dislocation subsystem does not lead to the same frustration, and
relaxation dynamics of this subsystem has a usual Arrenius form. It is $\beta$-relaxation.

This difference of the subsystems relaxation dynamics agrees with experimental results~\cite{9}, which prove that in
supercooled liquid there are two types of relaxation processes. These types concern the rotary and translational types
of diffusion. The degrees of freedom, which correspond to disclinations moves, are inherently rotational and correspond
to cooperative molecular motion, whereas the dislocation degrees of freedom are translational ones.

\section{Topology moment as an order parameter}

In some theories \cite{Rivier, Kiv1} the liquid--glass transition is considered as a special form of phase transition
in the glasses and structurally disordered systems. In this case the problem is reduced to description of the
interactive disclination subsystem. As in any another phase transition theory the central problem of this theory is the
definition of order parameter. On the one hand an order parameter should characterize the local state and describe the
short-range order symmetry, on the other hand the free energy of the system should be invariant under scaling
transformations, so that the scaling hypothesis was true.

The linear defect topology moment, $Q^{R}_{ij}$, used above satisfies these criterions. First of all this order
parameter characterizes the short-range order symmetry by definition. At the same time it has the non-molecular
(supermoleqular) scale, and the theory is scale-invariant. One can explain this in the following way: the local
ordering of the topological moments under decreasing of the temperature leads to the symmetry breakdown, SO(3)$\to
$SO(2). In this system the next generation of topology moments appears. At freezing in this system, just as in nematic,
an ordering of the moments occurs. However, as a result of their random situation a general direction for all
topological moments of the system does not exist. As a result this leads to the appearance of next generation of
topological defects with a bigger scale. After changing the scale one can describe the system with the former model.
Close to the critical point this succession is the scaling invariant hierarchy of the topology moments. This form of
order parameter was used in \cite{Vas} to describe the glass transition in terms of critical dynamics.

\section{Conclusions}

The important feature of the approach developed above is the possibility of the sequential description of supercooled
liquid. This description joins the macroscopic properties with the short-range and middle-range order of molecular
system. One can conclude that the presence of a local symmetry in a real molecular system leads to the breakdown of the
SO(3)-symmetry of the ideal continuous theory of defects. As a result the disclination and dislocation fields become
Abelian, and this essentially simplifies the system analysis. In the first infinitesimal order the free disclination
and dislocation fields are independent each other. Thus this approach is equivalent to the adiabatic approximation.
This also simplifies the system analysis but at the same time allows to keep the important properties of the system.
One of these properties is a decoupling between rotational and translational relaxation processes: according to the
developed theory $\alpha $-relaxation is the relaxation in the disclination subsystem, and $\beta $-relaxation is the
relaxation in the dislocation subsystem.

\section{Appendix 1 (topology)}

The formation of the bond orientational order in liquid leads to the addition of the U potential, fixing the local
orientation order symmetry, to the Lagrangian of the isotropic elastic matter. The state space of this system is the
set in which the free energy potential, U, extremes lie. It is the orbit of the U potential symmetry grope (the ground
state symmetry grope) on the symmetry grope, G, of the 3-d isotropic space, which coincides with the SO(3) rotation
grope.

The uniaxial nematic has the SO(2) ground state symmetry grope. Therefore its single orbit on the G is the sphere with
the identified antipode points, $\mathbb{R}$P$^2$=SO(3)/SO(2)$\times {\mathbb{Z}}^2$. The ground state symmetry grope
of icosahedral, Y, has the three orbits on the G, which are correspond to the discrete axial rotations around the 2-th,
3-th and 5-th order axes: SO(3)/Z$_{\nu_i}\times$Z$_2$ ($\nu_1=2, \nu_2=3, \nu_3=5$)~\cite{Nov}.

The initiation of a linear singularity (defect) in the system implies that from the ${\mathbb{R}}^3$ coordinate space
the $\mathbb{R}$ line is cut out. Thus the linear singularity presence leads to mapping of circle,
${\mathbb{R}}^3\backslash{\mathbb{R}}^1$$\sim$S$^1$, in the $M^n$ state space. If the class of the S$^1$ circle
homotopic mapping to the $M^n$ state space can not be contracted to point, $\pi _1(M^n)\neq $1, then the linear defect
in the order parameter field is stable. In both cases, which are of interest to us the linear singularities are
topologically stable, since $\pi_1$($\mathbb{R}$P$^2$)$\neq $1 and $\pi_1$(SO(3)/Y)$\neq $1. In the non-Abelian case,
when a symmetry breakdown is absent, there are no topologically stable defects, since $\pi_1$(SO(3)/SO(3))=0.

\section{Appendix 2 (the topology moment determination)}

In order to determine the disclination topology moment let us consider the tensor potential of an arbitrary system of
the closed linear disclinations, included in the volume, $V$:
\begin{equation}
\label{d5} A^i_a=e^i\dfr {\nu }2\,\varepsilon^i_{ak}\int \dfr{x^k}{x^3}l_i dV,
\end{equation}
where $x$ is the distance from the $l_idV$ disclination section to the observation point. Let us pick out an arbitrary
point in the considered system (<<center of currents>>), and signify by the $\vec R$ the vector from this point to the
disclination section ($\vec R=\vec x+\vec r$), then, provided that $r\ll x$, in the expansion (\ref{d5}) one can
content oneself with two first terms:
\begin{equation}
\label{d6}
\begin{array}{rl}
\displaystyle A^i_a&\displaystyle =e^i\dfr {\nu }2\,\varepsilon^i_{ak}\dfr{R^k}{R^3}\int l_i dV+e^i\dfr {\nu
}2\,\varepsilon^i_{ak}\int \dfr{r^k}{R^3}\,l_i dV+\\[12pt]
&\displaystyle +e^i\dfr {3\nu }2\,\varepsilon^i_{ak}\int \dfr{R^kR_sr^s}{R^5}\,l_i
dV+\ldots .
\end{array}
\end{equation}
In the order of values the relation of the last pair of terms to the first one is equal to $l/x$, where $l$ is the
linear dimension of $V$. Therefore at $x>l$, when this expression is correct, the last two terms are smaller than the
first one. Thus one can approximately write that
$$
A^i_a=e^i\dfr {\nu }2\,\varepsilon^i_{ak}\dfr{R^k}{R^3}\int l_i dV.
$$
However, in the case of the system of closed and screening each other topology defects, when
$$
\nu \int l_idV=0,
$$
the first and the second terms in (\ref{d6}) vanish, as a result
$$
\displaystyle A^i_a=\pi a^2\nu \,\left(\delta^i_a-3e^ie_a\right)\dfr{1}{R^3}+\ldots .
$$
In this case the topology moment, $Q^i_a=\pi a^2\nu \,\left(\delta^i_a-3e^ie_a\right)$, is the basic parameter, which
characterizes the closed linear defect system. At a great distance the tensor potential of this system is
$$
A^i_a=\dfr {Q^i_a}{R^3},
$$
where $R$ is the distance from the observation point to the disclination system, characterized by the topology moment
$Q^R$. Similarly one can express the tensor dislocation field potential by means of dislocation topology moment, $Q^T$.

\section{ACKNOWLEDGMENTS}
I am thankful to V.\,N.\,Rizhov for the helpful discussion of this paper. This work was supported by the RFBR grants
No. 07-02-00110-a and 07-02-96045-r$_{-}$ural$_{-}$a.


\section*{References}
\begin{thebibliography}{20}

\bibitem{Pat} Patashinski A Z and A. C. Mitus 1997 Towards understanding the local structure of liquids {\it Phys. Repts.} {\bf 288} 409

\bibitem{Tanaka} Tanaka H 2000 General view of a liquid--liquid phase transition {\it Phys. Rev. }E {\bf
62} 6968

\bibitem{4} Nelson D R 1983 Order, frustration, and defects in liquids and glasses {\it
Phys. Rev.} B {\bf 28}, 5515

\bibitem{Kov} Kovalenko K V, Krivokhizha S V, Chaban I A and Chaikov L L 2008 Detection of various phases in liquids from the hypersound velocity
and damping near closed phase-separation regions of solutions {\it JETP}   {\bf 106} 280

\bibitem{Stanley} Poole P H, Sciortino F, Essmann U and Stanley H E 1992 Phase behavior of metastable Water {\it Nature} {\bf 360} 324

\bibitem{1} Frencel Y I 1945 {\it Kinetic Theory of Liquid} (Moscow: AS USSR) p 424 (in Russian)

\bibitem{3} Tarjus G, Kivelson S A, Nussinov Z, Viot P 2005 The frustration-based approach of supercooled liquids and the glass
transition: a review and critical assessment {\it Journal of Physics: Condensed Matter} {\bf 17} R1143

\bibitem{Rivier} Rivier N, Duffy D M 1981 Line defects and the glass transition {\it Numerical methods in the study of
critical phenomena: Proc.Colloq. (Carry-le Rouet, France, June, 1980)} ed J Della Dora, J Demongeot, B Lacolle (Berlin:
Springer) pp. 132-142;

\bibitem{5} Likhachev V A, Volkov A E and Shudegov V E 1986 {\it Continual Theory of Defects}
(Leningrad: Leningrad university publishing house) p 228

\bibitem{6} Kadich A and Edelen D G B 1983 {\it A Gauge Theory of Dislocations and Disclinations}
(Berlin: Springer)

\bibitem{Vasin} Vasin M G, Lady'anov V I 2005 Description of fisher clusters formation
in supercooled liquids within framework of continual theory of defects {\it J. Phys.: Condens. Matter} {\bf 17} S1287

\bibitem{9} Debenedetti P G and Stillinger F H 2001 Supercooled liquids and the glass transition {\it Nature} {\bf 410}
259

\bibitem{Kiv1} Kivelson D, Tarjus G, Zhao X, Kivelson S A 1996 Fitting of viscosity: Distinguishing the temperature
dependences predicted by various models of supercooled liquids {\it Phys.Rev.E} {\bf 54} 5873

\bibitem{Vas} Vasin M G 2006 General approach to the description of the glass transition in terms of critical dynamics
{\it Phys.Rev.}B {\bf 74} 214116

\bibitem{Nov} Novikov S P, Taimanov I A 2006 {\it Modern Geometric Structures and Fields. Graduate Studies in Mathematics} {\bf
75} (AMS Providence: Road Island) p 633

\end{thebibliography}
\end{document}